%% file: HVP_paper_v3.tex
\documentclass[usenatbib,usegraphicx,useAMS]{mn2e}
\usepackage{amsmath}
\usepackage{amssymb}
\input{journal_abbr}  
\input{unit_abbr} 

\title{Hypervelocity Planets and Transits Around Hypervelocity Stars}

\author[Ginsburg, Loeb, Wegner]{Idan Ginsburg\thanks{E-mail:
idan.ginsburg@dartmouth.edu}, Abraham Loeb\thanks{E-mail:aloeb@cfa.harvard.edu} \& Gary A. Wegner\thanks{E-mail:gaw@dartmouth.edu} \\Department of Physics and 
Astronomy, Dartmouth College, 6127 Wilder Laboratory, Hanover, NH 03755, USA
\\Astronomy Department, Harvard University, 60 Garden St., Cambridge, MA 02138, USA\\}

\begin{document}
\maketitle

\begin{abstract}
The disruption of a binary star system by the massive black hole at
the Galactic Centre, SgrA*, can lead to the capture of one star around
SgrA* and the ejection of its companion as a hypervelocity star
(HVS). We consider the possibility that these stars may have planets
and study the dynamics of these planets. Using a direct $N$-body
integration code, we simulated a large number of different binary
orbits around SgrA*. For some orbital parameters, a planet is ejected
at a high speed. In other instances, a HVS is ejected with one or more
planets orbiting around it. In these cases, it may be possible to
observe the planet as it transits the face of the star. A planet may
also collide with its host star. In such cases the atmosphere of the
star will be enriched with metals. In other cases, a planet is
tidally disrupted by SgrA*, leading to a bright flare.
\end{abstract}


\section{Introduction} \label{INT}
Hypervelocity stars (HVSs) were first theorized in 1988
\citep{Hills:88}, and discovered observationally in 2005
\citep{Brown:05}. At least 16 HVSs have been identified in the Milky
Way (\citealt{Edelmann:05}; \citealt{Hirsch:05}; \citealt{Brown:06a};
\citealt{Brown:06b}; \citealt{Brown:07b}; \citealt{Brown:09b}).There
are a number of proposed mechanisms for the production of HVSs. The
best studied and arguably most likely mechanism is the Hills' scenario
where the close interaction of a binary star system and massive black hole
(MBH) can produce a HVS with sufficient velocity to escape the
gravitational pull of the Milky Way
(\citealt{Hills:88,Gould-Quillen:03}; \citealt{Yu-Tremaine:03};
\citealt{Ginsburg:1}; \citealt{Perets:07}; \citealt{Perets:09a};
\citealt{Madigan:09,Madigan:11}). Other mechanisms include the
interaction of stars with stellar black holes \citep{Oleary-Loeb:08},
the inspiral of an intermediate mass black hole 
(\citealt{Hansen:03}; \citealt{Yu-Tremaine:03}; \citealt{Levin:06};
\citealt{Sesana:09}), and the disruption of a triple system
(\citealt{Lu:07}; \citealt{Sesana:09}; \citealt{Perets:09b};
\citealt{Ginsburg:3}). We focus on the disruption of a tightly bound
binary by a central black hole of inferred mass $\sim 4 \times 10^6
M_{\odot}$ (e.g. \citealt{Scho:03}; \citealt{Reid-Brunthaler:04};
\citealt{Ghez:05}; \citealt{Ghez:08}; \citealt{Gillessen:2009}).

Simulations show that tight binaries disrupted by the MBH can result 
in the ejection of one component with velocities
comparable to those of the observed HVSs
(e.g. \citealt{Ginsburg:1,Bromley:06}), and perhaps significantly
larger velocities as well \citep{Sari:10}. When a hypervelocity star
is produced, the companion to the HVS is left in a highly eccentric
orbit around SgrA* \citep{Ginsburg:1}. Furthermore, a small fraction
of binary systems may collide when disrupted, and if the collision
velocity is small enough the system can coalesce
\citep{Ginsburg:2,Antonini:11}. For planets orbiting a star, tidal
dissipation will eventually result in the infall of short-period
planets into their host star (e.g. \citealt{Rasio:96,Jackson:09});
however the timescale is on the order of Gyrs, and planets with
orbital periods of $\lesssim 1$ day are predicted to survive while the
host star is on the main-sequence \citep{Hansen:10}. A number of
Jupiter-mass planets have already been found with orbital periods
$\lesssim 1$ day (e.g. \citealt{Hebb:10,Hellier:11}). 
Thus, it is only natural to consider the scenario where
planets are orbiting such tightly bound binary systems which are
subsequently disrupted by the MBH. Our goal is to examine the possible
orbital dynamics and determine whether a HVS may host a planetary
system.

In \S 2 we describe the codes and simulation parameters. In \S 3 we
discuss the origin of hypervelocity planets (HVPs) and transits around HVSs, and in \S 4 we
discuss some possible outcomes for the disruption of a binary system
with planets. Our goal was not to cover the entire available phase
space, but to determine whether some tight binaries with planets could
produce transits or other observable effects.

\section{Computational Method} \label{CM} 
In our study we have used the publically available N-body code written
by \citet{Aarseth:99}
\footnote{http://www.ast.cam.ac.uk/$\sim$sverre/web/pages/nbody.htm}.
We adopted a small value of $10^{-8}$ for the accuracy parameter
$\eta$, which determines the integration step through the relation
$dt=\sqrt{{\eta F}/(d^2F/dt^2)}$ where $dt$ is the timestep and $F$ is
the force.  The softening parameter, \emph{eps2}, which is used to
create the softened point-mass potential, was set to zero. We treat
the stars and planets as point particles and ignore tidal and general
relativistic effects on their orbits, since these effects are small at
the distance ($\gtrsim10$AU) where the binary star system is tidally
disrupted by the MBH. In total, we ran over 8000 simulations.

We set the mass of the MBH to $M=4\times 10^6M_{\odot}$, and the mass
of each star to $m=3M_{\odot}$, which is comparable to the known
masses of HVSs (\citealt{Fuentes:06,Przybilla:08b}). Since the
planet's mass $\ll m_{\star}$ we treat all planets as test
particles. All runs start with the centre of the binary system located 2000
AU ($=10^{-2}$pc) away from the MBH along the positive y-axis. This
distance is larger than the binary size or the distance of closest
approach necessary to obtain the relevant ejection velocity of HVSs,
making the simulated orbits nearly parabolic. Our simulations include
orbits in a single plane as well as simulations with the binary coming
out of the orbital plane at $90^\circ$. For simplicity, the planetary
orbital plane was kept the same as that of the binary. We used the
same initial distance for all runs to make the comparison easier to
interpret as we varied the distance of closest approach to the MBH or
the relative positions of the stars and planets within the system.

We ran two primary sets of simulations. The first set consisted of a
binary system with two planets. Each star had one planet initially at
$a_p = 0.02$ AU from its host star on a circular orbit (with
eccentricity $e=0$). The second set consisted of a binary system with
four planets. Initial conditions were similar to the previous data
set, with the exception that the second planet was placed on a circular
orbit with distance $a_p = 0.03$ AU from the star. We varied the
initial separation between the two stars from $a_{\star} = 0.05$ to
$0.5$ AU and precluded tighter binaries for which two stars develop a
common envelope and coalesce. Similarly, much wider binaries may not
produce HVSs (\citealt{Hills:1991,Bromley:06,Ginsburg:1}). However, in our
simulations we assume planetary distances significantly shorter than
$0.02$ AU would bring the planet into the star's atmosphere and thus
are precluded. Similarly, too large a distance could result in a
dynamical instability whereby the planets are ejected from the system
or collide with a star. In order to determine the effect of the
planetary separation on the orbital dynamics, we ran additional
simulations with a fixed binary semimajor axis $a_{\star} = 0.2$ AU
and varied the planetary separation in the range $a_p = 0.02$--$0.06$
AU. Our simulations excluded planets around single stars since such systems 
will never produce HVSs.

A three-body encounter can lead to chaos, and the initial phase of the
stellar orbit can greatly vary the outcome
\citep{Ginsburg:1}. Therefore, we sampled cases with initial phase
values of $0$--$360^o$ at increments of $15^o$.  We gave the binary
system no radial velocity but a tangential velocity with an amplitude
such that the effective impact parameter ($b$) is in the range of
$5$--$35$ AU.  The distribution of impact parameters is as follows: our
simulations had $b = 5, 10, 15, 20$ AU with a likelihood of $\sim
20$\% of all runs for each value, while the remainder at $b = 25$ or
$30$ AU had a $\sim 10\%$ likelihood each.
We expect no HVSs to be produced at substantially larger impact
parameters \citep{Ginsburg:1}.

\section{Origin of Hypervelocity Stars and Hypervelocity Planets} \label{HVP}

Given a binary system with stars of equal mass $m$ separated by a
distance $a$ and a massive black hole (MBH) of mass $M\gg m$ at a
distance $b$ from the binary, tidal disruption would occur if $b\la
b_{\rm t}$ where
\begin{equation}
\frac{m}{a^3} \sim \frac{M}{{b^3_{\rm t}}}.
\end{equation}
The distance of closest approach in the initial plunge of the binary
towards the MBH can be obtained by angular momentum conservation from
its initial transverse speed $v_{\perp}$ at its initial distance from
the MBH, $d$,
\begin{equation}
v_{\perp}d = \left(\frac{GM}{b}\right)^{1/2}b .
\end{equation}
The binary will be tidally disrupted if its initial transverse speed
is lower than some critical value,
\begin{equation}
v_\perp\la v_{\perp,\rm crit} \equiv {(GMa)^{1/2}\over d}\left({M\over m}\right)^{1/6}= 10^2 {a_{-1}^{1/2} \over m_{0.5}^{1/6} d_{3.3}} ~{\rm {km~s^{-1}}},
\label{eq:crit}
\end{equation}
where $a_{-1}\equiv ({a}/{0.1~{\rm AU}})$, $d_{3.3}=(d/2000~{\rm
AU})$, $m_{0.5} \equiv (m/3M_{\odot})$, and we have adopted $M=4\times
10^6M_{\odot}$.  For $v_\perp\la v_{\perp,\rm crit}$, one of the stars
receives sufficient kinetic energy to become unbound, while the second
star is kicked into a tighter orbit around the MBH.  The ejection
speed, $v_{\rm ej}$, of the unbound star can be obtained by
considering the change in its kinetic energy $\sim v\delta v$ as it
acquires a velocity shift of order the binary orbital speed $\delta v
\sim \sqrt{Gm/a}$ during the disruption process of the binary at a
distance $\sim b_t$ from the MBH when the binary centre-of-mass speed
is $v\sim \sqrt{GM/b_t}$ \citep{Hills:88,Yu-Tremaine:03}. At later
times, the binary stars separate and move independently relative to
the MBH, each with its own orbital energy.  For $v\la v_{\perp,\rm
crit}$, we therefore expect
\begin{align}
v_{\rm ej} \sim
\left[\left({\frac{Gm}{a}}\right)^{1/2}\left({\frac{GM}{b_{\rm
t}}}\right)^{1/2}\right]^{1/2} \nonumber\\ = 1.7 \times 10^3
m^{1/3}_{0.5}a^{-1/2}_{-1} ~{\rm km~s^{-1}}.
\label{eq:model}
\end{align}

\begin{figure}
\begin{center}
\includegraphics[width=0.5\textwidth]{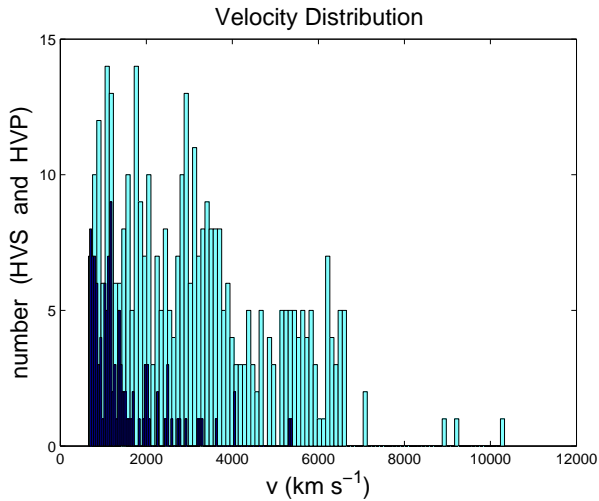}
\end{center}
\caption{
Velocity distribution of HVSs and HVPs. This sample comes from 1000
simulations. The initial binary separation for each system was
$a_{\star}$ = 0.2 AU, and we let the planetary separation vary with
uniform probability in the range $a_p = 0.02$--$0.04$ AU. These runs
are strictly for systems with two planets, however we get similar
distributions for systems with four planets. The lowest HVP velocities
are $\sim 700$ km s$^{-1}$, which corresponds with the lowest
velocities for HVSs. The average HVP velocity is $\sim 3000$ km
s$^{-1}$, and the average HVS velocity is $\sim 1500$ km s$^{-1}$ (see
also Table \ref{tab_v}). The HVPs are denoted in light blue, and the
HVSs in the darker color. The overall shape of the distribution is
similar in both cases, and in both cases there are outliers. Note that
there are $\sim 3$ HVPs for each HVS.}
\label{fig:hist}
\end{figure}

A planet with mass $m_p \ll m_{\star}$ may also be ejected at even
higher speeds. The mechanism that ejects HVPs involves the interaction
of the MBH, the binary star system, and the planet, and does not admit
a simple analytical estimate.  Based on our simulations, the
velocities of hypervelocity planets (HVPs) are on average $\sim 1.5$--$4$
times the velocity of a typical HVS (see Table \ref{tab_v}).
We found a few examples of HVPs with exceptionally high speeds, 
$v\sim10^4$ km s$^{-1}$,
however as observed in Figure \ref{fig:hist}, these are rare with a
probability $< 1$\%.  Although Figure \ref{fig:hist} represents one
specific parameter set, we do get comparable results for other runs.

\begin{table}
\begin{center}
\begin{tabular}{|r|r|r|r|r|r|}
\hline
$a_{\star}$ (AU)&$\bar{v}_{HVS}$ (km~s$^{-1}$) &$\bar{v}_{HVP}$ (km~s$^{-1}$)\\
\hline
0.05&2700&3800\\
0.10&2000&3100\\
0.20&1400&3500\\
0.30&1300&4100\\
0.40&1400&4100\\
0.50&1100&4400\\
\hline
0.05&-&-\\
0.10&-&-\\
0.20&1500&3300\\
0.30&1300&3900\\
0.40&1400&4200\\
0.50&1100&4200\\
\hline
\end{tabular}
\end{center}
\caption{Average velocity of HVSs (second column) and HVPs (third
column) for different values of $a_{\star}$. The top four rows show
the values obtained from our simulations with two planets, and the
bottom rows our results with four planets.  For our simulations
with four planets, the outer planets either collided with a star or were
immediately ejected when $a_{\star} < 0.2$ AU, hence no values are
listed for 0.05 and 0.10 AU. All values are uncertain to within $\pm
50$ km~s$^{-1}$}.
\label{tab_v}
\end{table}

\begin{figure*}
\begin{center}
\begin{tabular}{ccc}
&
\includegraphics[width=0.5\textwidth]{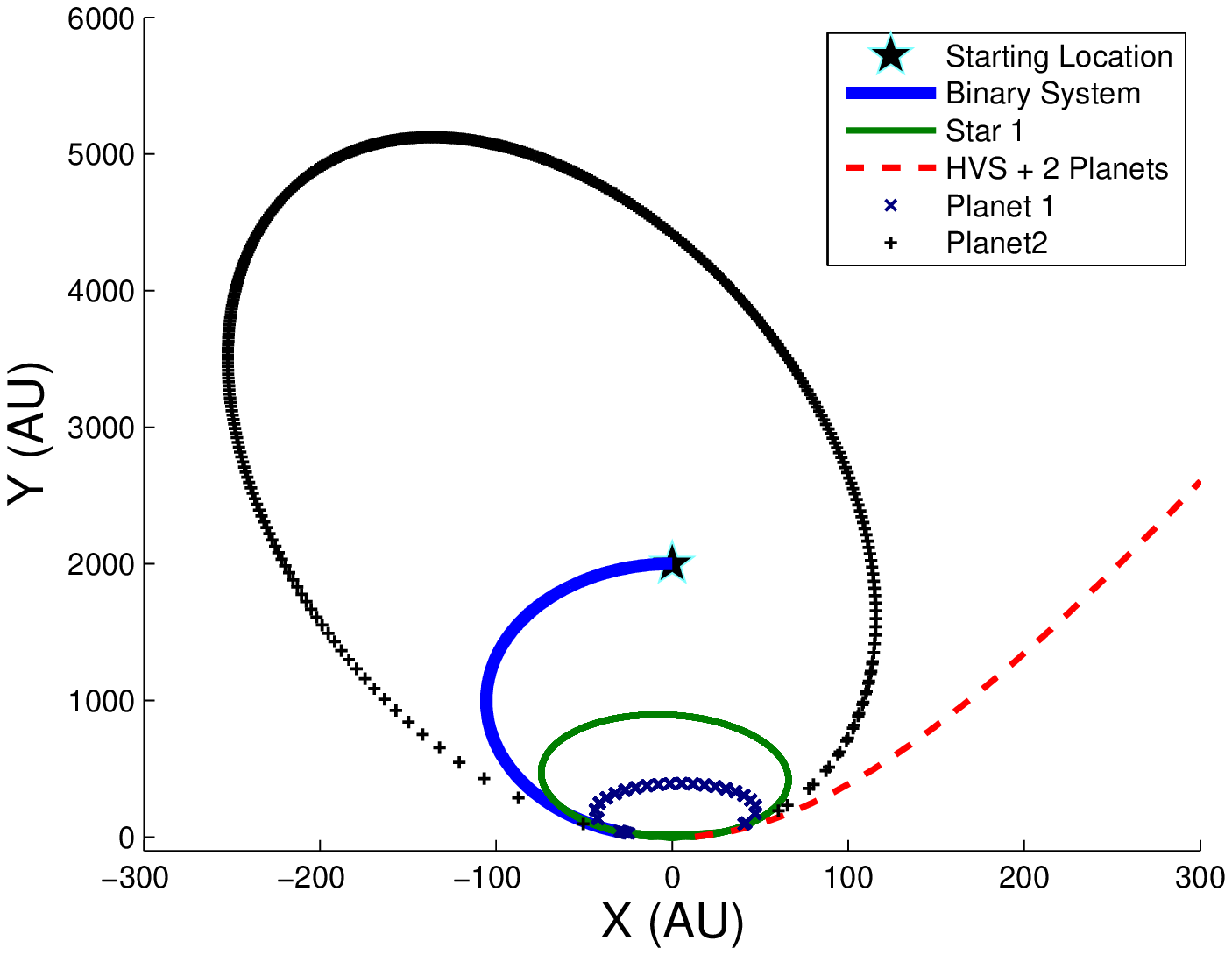}
&
\includegraphics[width=0.5\textwidth]{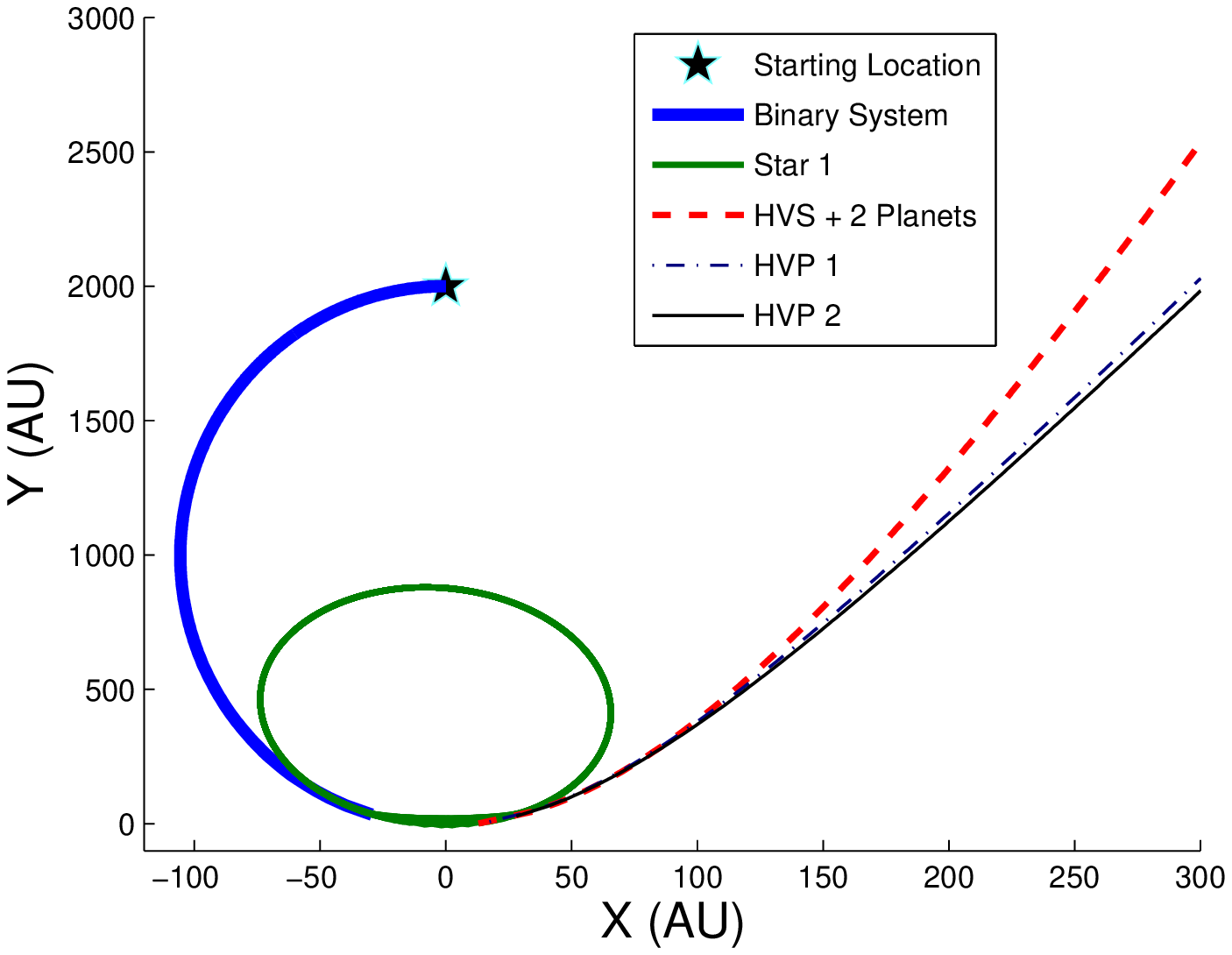}
\\
\end{tabular}
\end{center}
\caption{The two panels illustrate possible outcomes after a binary
system with planets is disrupted by the MBH. The MBH is located at the
origin, and the binary system starts at an initial distance of 2000 AU
along the positive y-axis. In both cases the initial binary separation
was $a_{\star} = 0.2$ AU and the planetary separation $a_p = 0.02$ AU
for the innermost planets and $a_p = 0.03$ AU for the outer
planets. {\bf Left:} After binary disruption a HVS is produced with
two bound planets, as marked by the dashed line. The second star,
marked by the solid line, stays in a highly eccentric orbit around the
MBH. The second star's planets are removed, and the first falls into a
highly eccentric orbit close to the MBH, while the second is ejected
into a much larger, but also highly eccentric orbit around the
MBH. {\bf Right:} After binary disruption a HVS is produced with two
planets in orbit as marked by the dashed line. The second star, marked
by the solid line, remains in a highly eccentric orbit around the
MBH. The second star's planets are ejected as HVPs.}
\label{fig:orbits}
\end{figure*}

\section{The Fate of Planetary Systems}
\label{outcomes}

We analyze statistically the orbital properties of the binary
star system after it is disrupted by the MBH. We ran simulations both
planar and with the binary out of the orbital plane at $90^\circ$. In our
simulations the initial planetary semimajor axis is constrained. Too
small a separation would lead to a plunge of the planet into the host
star's atmosphere, and too large a separation would result in an
unstable configuration in which the planet ultimately collides with a
star or else is ejected from the system. 

Figure \ref{fig:orbits} illustrates two possible outcomes. In both
instances, HVSs with orbiting planets are produced (dashed line),
while the companion star (solid line) orbits the MBH in a highly
eccentric orbit. The companion star's planets are themselves removed,
and either stay bound to the MBH in different orbits or are ejected as
HVPs.  Table \ref{tab_prob} shows various outcomes from our
simulations including the fraction of HVPs, planets around HVSs, and
free planets around the MBH.

\begin{table*}
\begin{center}
\begin{tabular}{|r|r|r|r|r|r|}
\hline
$a_{\star}$ (AU)& HVSs & HVPs & HVSs + Planets &Bound Stars + Planets&Free Planets\\
\hline
0.05&0.40$\pm0.03$ &0.36$\pm0.03$&0.05$\pm0.01$&0.08$\pm0.01$&0.80$\pm0.03$\\
0.10&0.37$\pm0.03$ &0.36$\pm0.03$&0.08$\pm0.01$&0.35$\pm0.03$&0.63$\pm0.03$\\
0.20&0.20$\pm0.02$ &0.42$\pm0.03$&0.05$\pm0.01$&0.40$\pm0.03$&0.58$\pm0.02$\\
0.30&0.07$\pm0.01$ &0.39$\pm0.03$&0.02$\pm0.01$&0.35$\pm0.03$&0.62$\pm0.03$\\
0.40&0.03$\pm0.01$ &0.35$\pm0.03$&0.01$\pm0.01$&0.39$\pm0.03$&0.62$\pm0.03$\\
0.50&0.02$\pm0.01$ &0.43$\pm0.03$&0.01$\pm0.01$&0.42$\pm0.03$&0.57$\pm0.02$\\
\hline
0.05&-&-&-&-&-\\
0.10&-&-&-&-&-\\
0.20&0.19$\pm0.02$&0.83$\pm0.04$&0.11$\pm0.02$&0.59$\pm0.04$&0.64$\pm0.02$\\
0.30&0.05$\pm0.01$&0.70$\pm0.04$&0.03$\pm0.01$&0.59$\pm0.04$&0.67$\pm0.02$\\
0.40&0.02$\pm0.01$&0.71$\pm0.04$&0.01$\pm0.01$&0.63$\pm0.04$&0.65$\pm0.02$\\
0.50&0.02$\pm0.01$&0.78$\pm0.04$&0.01$\pm0.01$&0.76$\pm0.04$&0.60$\pm0.02$\\
\hline
\end{tabular}
\end{center}
\caption{
Probability for various outcomes in our ensemble of runs. 
The top rows show the results for simulations with two planets, and the 
bottom rows for our simulations with four planets. Note that for our
simulations with four planets, the outer planets either collided with a star
or were immediately ejected when $a_{\star} < 0.2$ AU, hence no values 
are listed for 0.05 and 0.10 AU. The distance between each planet and host 
star is $a_p = 0.02$ AU, and for simulations with four planets the 
second set of planets had $a_p = 0.03$ AU. 
The first column is the initial distance between the two stars. 
The second column shows the fraction of 
HVSs produced, and as expected tighter binaries produce more HVSs. The third column shows
the rate of HVP production. 
The fourth column shows the probability of producing HVSs with planets 
in orbit. This column is also represented graphically in the middle panel of
Figure \ref{fig:prob_HVS}. The fifth column shows the fraction of stars that are bound to the
MBH and have planets in orbit which may produce transits (see Section \ref{Tidal}). 
The last column shows the fraction of free planets around the MBH. Probabilities in each row are 
independent of each other. Note that the sum of the probabilities for each row add
to more than one due to the fact that we have two or four stars for each simulation. 
Therefore, as illustrated in Figure \ref{fig:orbits}, there may be multiple outcomes
for any given simulation. The quoted values include Poisson errors. }
\label{tab_prob}
\end{table*}

\subsection{HVPs} \label{HVP}

Our simulations show that for tight binaries with semimajor axes in
the range $a_{\star} = 0.05$--$0.5$ AU and planets with semimajor axes
$a_p = 0.2$--$0.05$ AU, the probability of producing HVPs is high (see Table \ref{tab_prob}).
For a binary with two planets, we find that the probability of
producing a HVP is $\sim 30$--$40$\%. For a binary with four planets,
the probability rises to $\sim 70$--$80$\%. However, since planets are
extremely faint, it is impossible to detect HVPs directly with
existing telescopes.

\subsection{Transits Around HVSs} \label{Transit}

The probability of observing an eclipse as a planet orbits its
host star is the transit probability ($P_T$) given by
\begin{align}
P_{T} = \left(\frac{R_{\star}\pm R_P}{a}\right)\left(\frac{1+e
\sin\omega}{1-e^2}\right),
\label{eq:transit}
\end{align}
where $R_{\star}$ is the star's radius, $R_P$ the planet's radius, $e$
is the eccentricity, and $\omega$ is the argument of periapse (see
\citealt{Murray-Correia}; \citealt{Winn}).  Note that the $+$ sign in
equation (\ref{eq:transit}) includes grazing eclipses, which are
otherwise excluded.  When $R_{\star} \gg R_P$ and $e=0$, we get
\begin{align}
P_{T} \approx
0.015\left(\frac{R_{\star}}{3R_{\odot}}\right)\left(\frac{1 {\rm
AU}}{a}\right).
\label{eq:transit2}
\end{align}
For massive ($m\geq1M_{\odot}$) stars on the main-sequence, it is 
safe to assume $R_{\star} \gg R_P$ and thus the probability of a 
transit is independent of planetary radius. However, the fractional
flux decrement is given by $(R_p/R_{\star})^2$. Therefore a larger
planetary radius is important in order to be able to get proper
photometry. We assume a Jupiter-sized planet, thus $(R_p/R_{\star})^2 \sim 1$\%.
Our simulations produced planets orbiting HVSs with $\bar e$ = 0.6. 
Although an eccentricity will increase the probability of a transit, 
equation (\ref{eq:transit}) is sensitive to high eccentricity and for a 
limiting case it is reasonable to use $e=0$. The timescale for eccentricity damping,
whereby a planet's eccentricity is circularized is not well
understood, and depending on initial conditions one can get either
short ($\sim 100$ Myr) or long ($\sim$ Gyr) circularization times
\citep{Naoz2}. Furthermore,  $e=0$ is consistent with the observed 
eccentricities for most extrasolar planets with $a_p < 0.1$ AU
(e.g. \citealt{Jackson:08,Hansen:10,Matsumura:10}).

\begin{figure*}
\begin{center}
\begin{tabular}{ccc}
\includegraphics[width=0.35\textwidth]{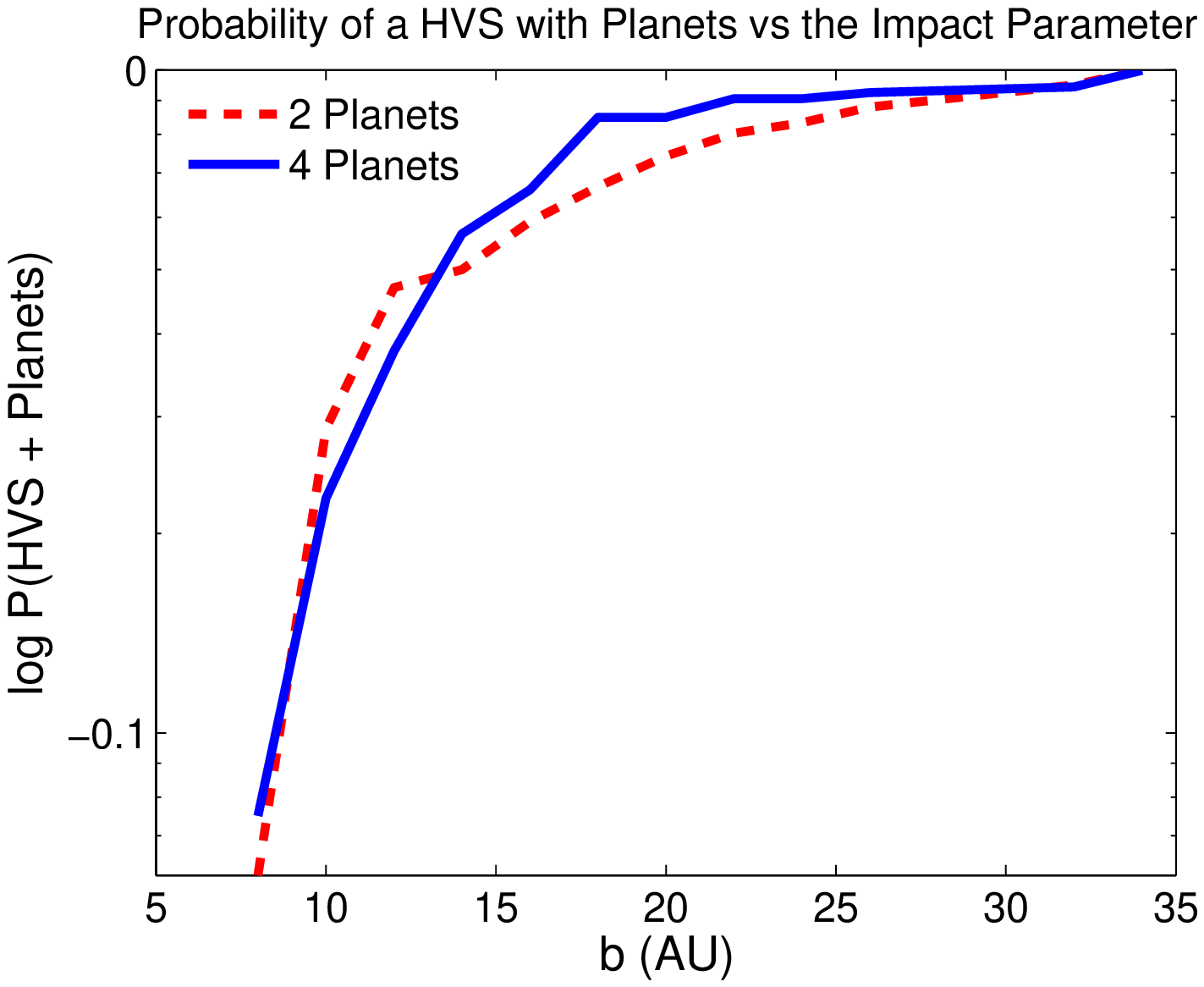}
&
\includegraphics[width=0.35\textwidth]{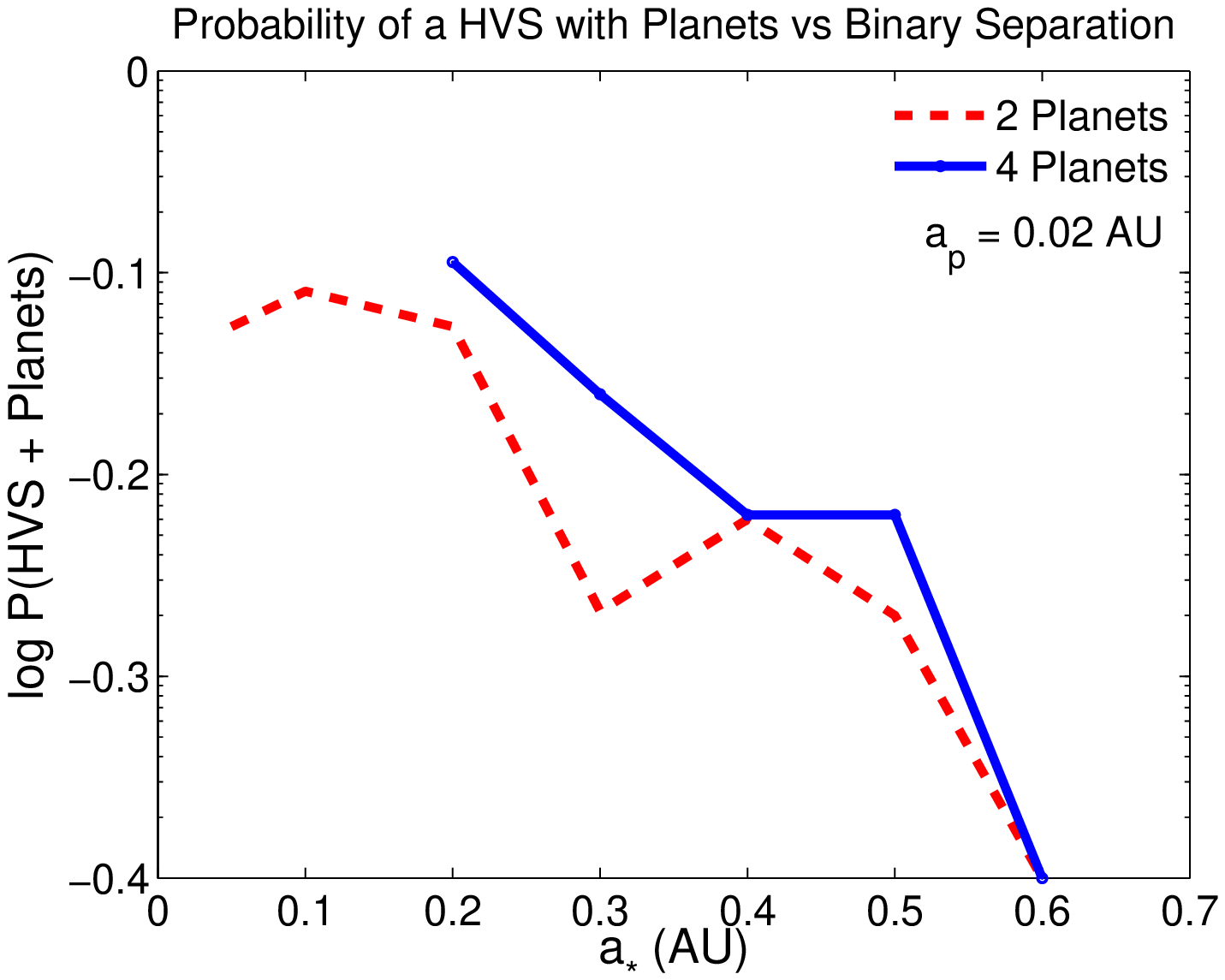}
&
\includegraphics[width=0.35\textwidth]{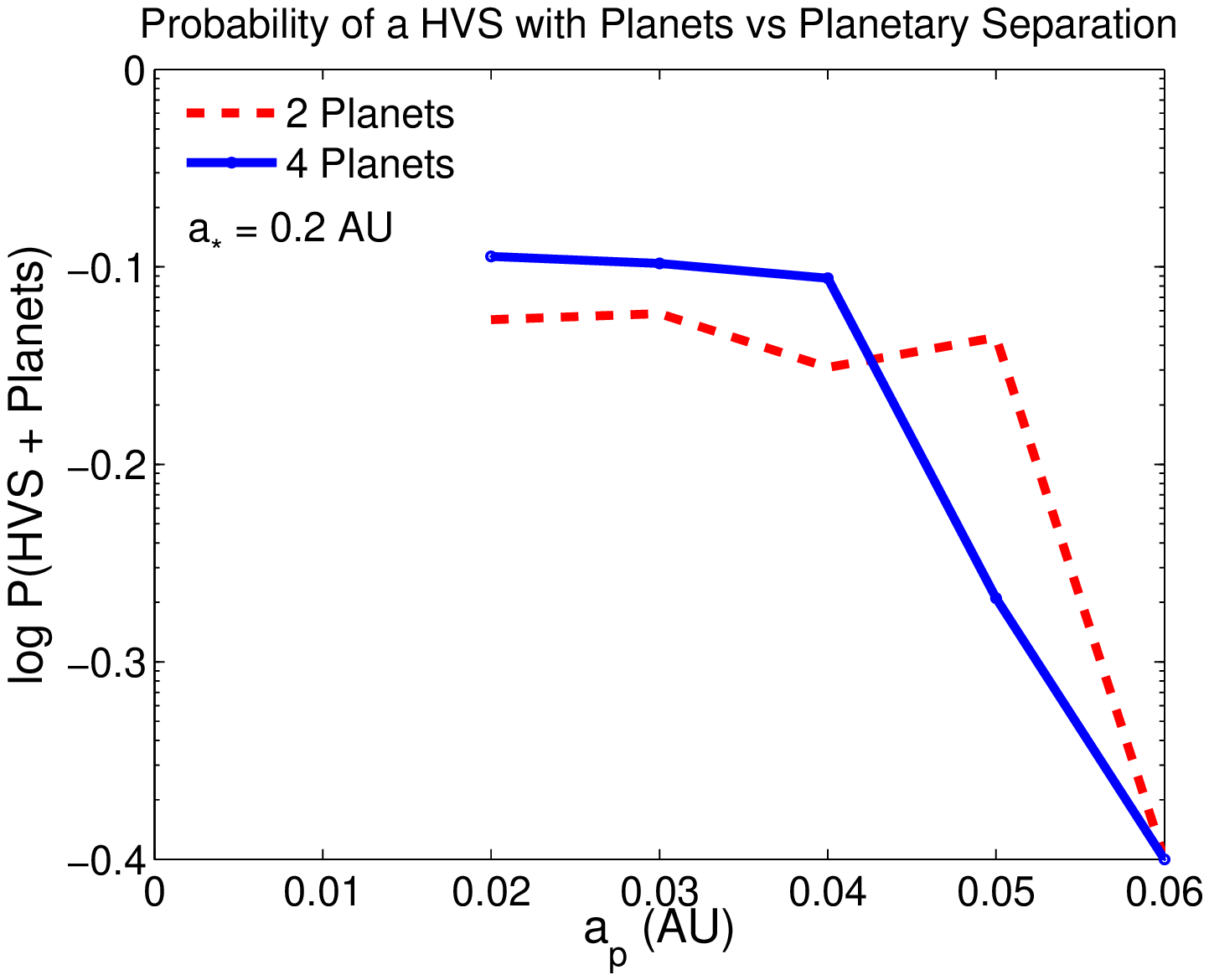}
\\
\end{tabular}
\end{center}
\caption{
Probability of producing a HVS with orbiting planets versus
various parameters. {\bf Left:} Probability as a function of impact
parameter, $b$, relative to SgrA*. This cumulative plot shows that an
impact parameter between $10-20$ AU is optimal.  The probabilities
here are defined relative to all our HVS runs.  {\bf Middle:}
Probability as a function of the initial distance between the two
stars. In this scenario we exclude simulations with four planets when
$a_{\star} < 0.2$ AU, due to the fact that at such distances the two
outermost planets were disrupted and either collided with a star or
were ejected. The initial planetary separation is $a_p = 0.02$ AU for 
all runs with two planets, and for our simulations with four planets the 
second set of planets had initial distance $a_p = 0.03$ AU.
{\bf Right:} Probability as a function of initial planetary
separation, $a_p$. We precluded orbits with $a_p<0.02$ AU, since such
orbits would take the planet into the star's atmosphere. The results
for each figure are averaged for all orbits, those in a single plane
and orbits out of the plane. The values of $a_p$ displayed along the
$x$-axis correspond to the first set of planets. For simulations with 
four planets, the second set was placed $0.01$ AU from the first.
Note that for the middle and right panels the impact parameter is 
not constrained.}
\label{fig:prob_HVS}
\end{figure*}

\begin{figure*}
\begin{center}
\begin{tabular}{ccc}
\includegraphics[width=0.5\textwidth]{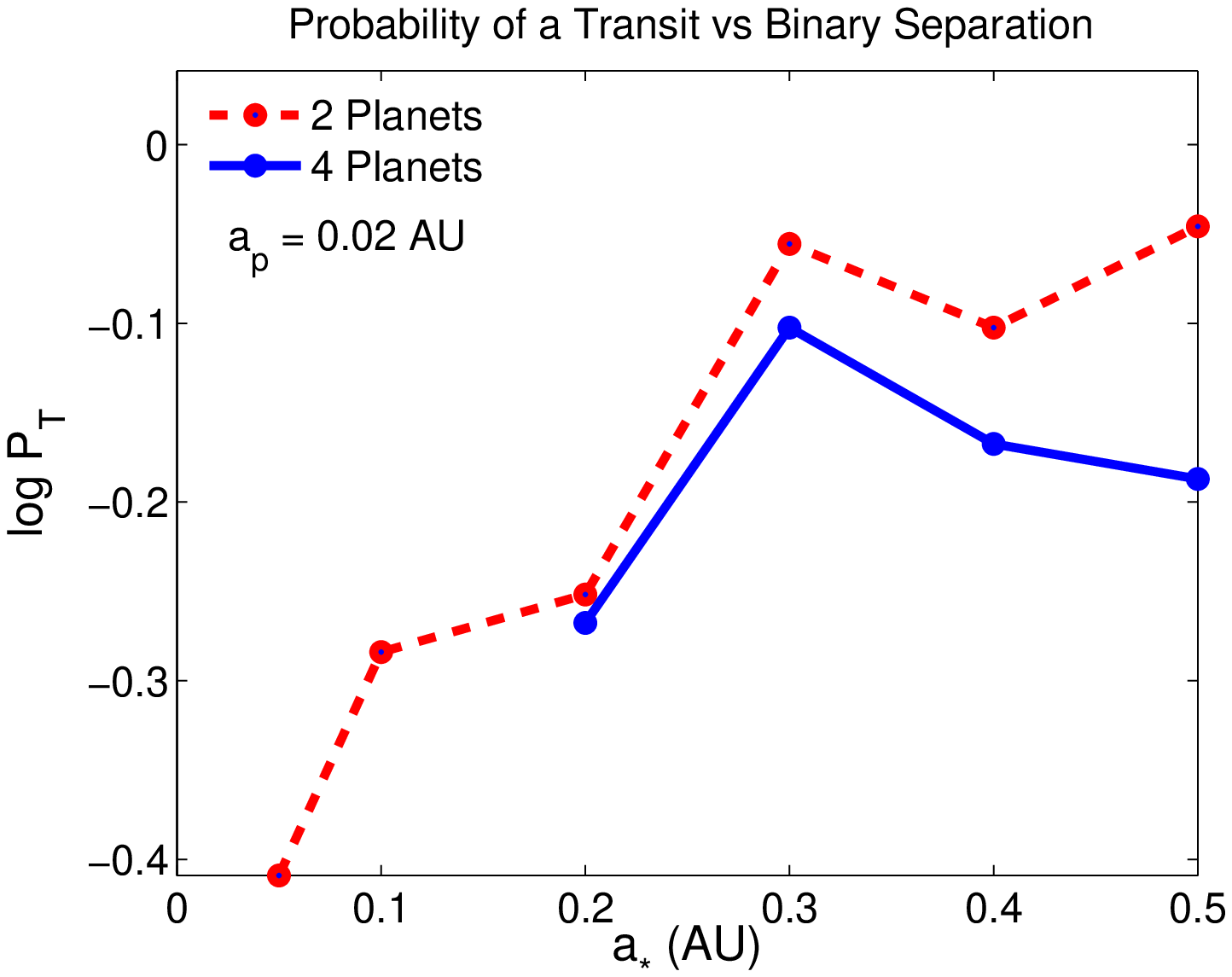}
&
\includegraphics[width=0.5\textwidth]{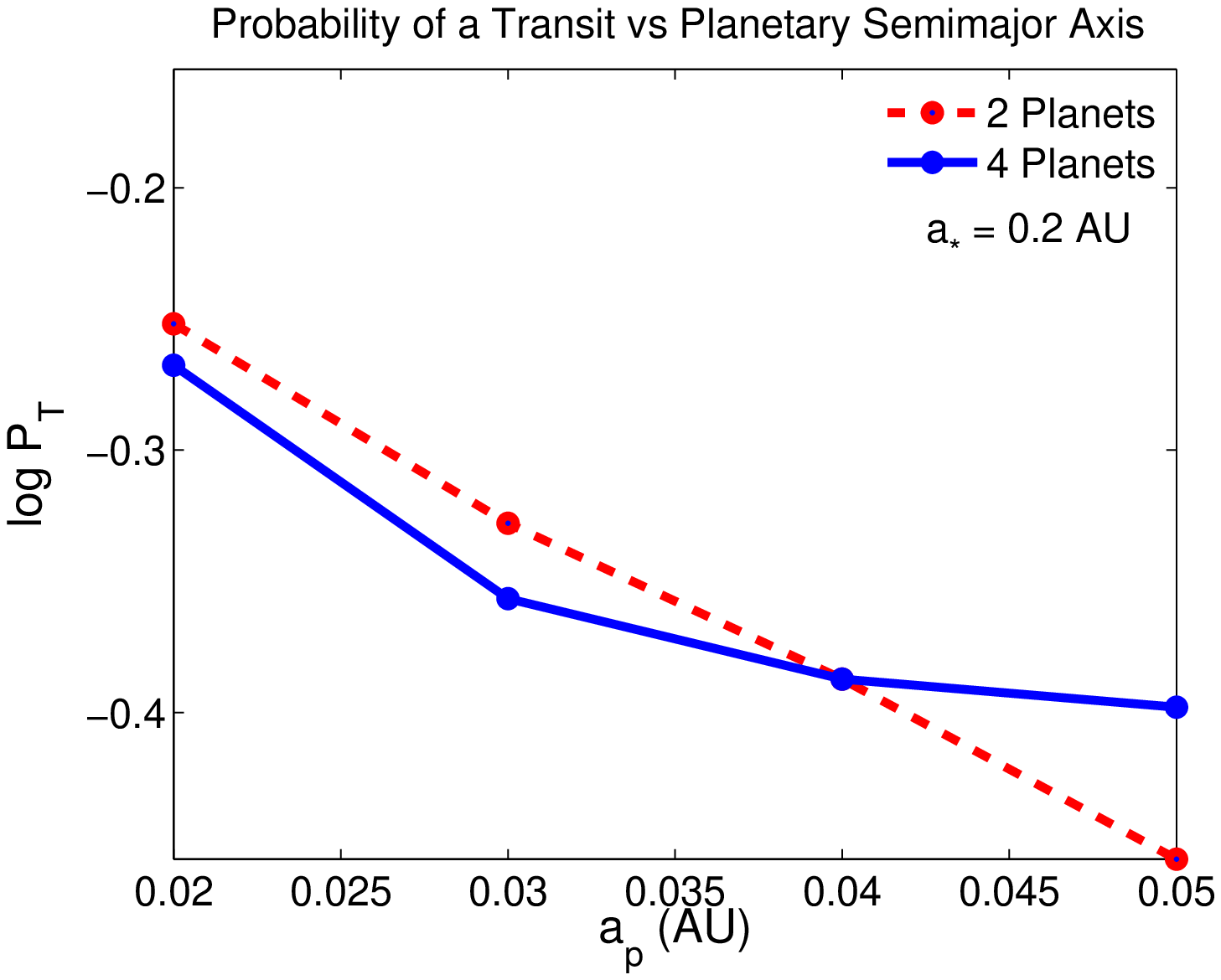}
\\
\end{tabular}
\end{center}
\caption{Probability of observing a transit for a planet orbiting a
HVS. {\bf Left:} Probability of a transit (log $P_T$) vs the initial
binary separation ($a_{\star}$). We exclude simulations with four
planets when $a_{\star}<0.2$ AU, since the outermost planets are
unstable at these parameters. {\bf Right:} Probability of a transit (log $P_T$) vs
the initial distance between the planet and host star ($a_p$). In this
scenario $a_{\star}$ is fixed at 0.2 AU. In both instances, our
results for two planets and four planets are qualitatively similar. The
probability of observing a transit ranges from $0.35-0.90$.}
\label{fig:transit}
\end{figure*}

Table \ref{tab_prob} shows the probability of producing a HVS with
orbiting planets.  This probability is calculated as the fraction of
all the runs we conducted in the restricted range of parameters for
which HVSs are produced.  We used a binary semimajor axis between
$a_{\star}=0.05$--$0.5$ AU. Smaller $a_{\star}$ would lead to a common
envelope and are precluded.  Larger $a_{\star}$ are likely to either
eject the planet from the system before producing a HVS, lead to a
collision, or fail to produce a HVS altogether.  The results for our
simulations with two planets and four planets are qualitatively
similar with some expected statistical variations, as illustrated in
the middle panel of Figure \ref{fig:prob_HVS}.  Our simulations also
indicate that the planetary semimajor axis needs to be between $a_p =
0.02$--$0.05$ AU in order for a planet to remain bound to a
HVS. Smaller $a_p$ are precluded due to the fact that they would sink
into the star's atmosphere, and at larger $a_p$ the planets are
tidally removed before the HVS is produced.  For two-planet systems
with $a_{\star} = 0.2$ AU, we find that when $a_p = 0.02$--$0.05$ AU
the probability of producing a HVS with planets is constant. We get
similar results for four planets, however the probability starts to
rapidly decrease when $a_p = 0.04$ AU.  These results are illustrated
in the right panel of Figure \ref{fig:prob_HVS}.  In total, we find
that the probability of producing a HVS with planets peaks at $\sim
10$\%.  Assuming a HVS has planets in orbit, the probability of
observing a transit is given by equation (\ref{eq:transit2}). We plot
these probabilities in Figure \ref{fig:transit}.

\subsection{Collisions} \label{Coll}

Assuming that a binary is disrupted at a random angle and ignoring
gravitational focusing, the probability for a collision is
\begin{equation}
P \sim \frac{4R_*}{2\pi R_{\rm sep}} ,
\end{equation}
where $R_{\star}$ is the radius of the star (assuming radii are
approximately equal) and $R_{\rm sep}$ is the average separation of the
stars. The stars will merge if the relative velocity of impact is less
than the escape velocity from the surface of the star ($\sim 500 ~{\rm
km~s}^{-1}$). A simple estimation for the impact velocity comes from
conservation of energy,
\begin{equation}
E = \frac{1}{2}\frac{m_1m_2}{m_1+m_2}\dot r^2 - \frac{Gm_1m_2}{r} =
const ,
\end{equation}
which yields the relative impact velocity for two stars
\begin{align}
v_f = \left[2G(m_1+m_2)\left(\frac{1}{a_{min}} -
\frac{1}{a}\right)\right]^{1/2}.
\label{energy}
\end{align}
Similarly, for a planet around a star, we arrive at
\begin{align}
v_f \sim \left[Gm_{\star}\left(\frac{2a-r}{ra}\right)\right]^{1/2}.
\label{energy_p}
\end{align}
Equations (\ref{energy}) and (\ref{energy_p}) yield a velocity similar
to the escape speed from the surface of the star $v \sim500$ km~${\rm
s}^{-1}$, however they do not take tidal forces nor gravitational
focusing into account, therefore the actual impact velocities are
expected to be slightly higher.  For a Jupiter-like planet, a direct
collision will release at minimun $10^{45}$ erg, and thus might be
observable as a flare.

The rate of HVS production in the
Milky Way is $\sim 10^{-5} {\rm ~yr}^{-1}$ \citep{Brown:06b}, and the
total collisional rate between stars is $\sim 10$ \%
\citep{Ginsburg:2}. Based on our simulations, we find that the collisional 
probability for a system with two planets is $\sim 0.1$\%, whereas 
for a system with four planets it is $\sim 1$\%. For systems with four 
planets, the collisions are almost entirely due to the two outer planets.

Should a planet collide with its host star, the planet will be
destroyed and add high-metallicity material to the star's surface. The
metallicity of stars with planetary systems has been shown to
be greater than expected by a factor of $\sim 2$
(e.g. \citealt{Santos:03,Vauclair}). Such high metallicity may be due
to tidal interactions that have caused exoplanets to fall into their
host stars \citep{Levrard:09,Jackson:09}.  Although the mechanism by
which a star may lose its high-metallicity material is not well
understood \citep{Sasselov}, \citet{Garaud} has shown that the
relative metallicity enhancement on the surface due to the infall of
exoplanets, survives a much shorter period for higher-mass stars. A
star of $1.5M_{\odot}$ would lose $\sim 90$\% of its enhanced
metallicity in merely 6 Myr. Thus, a HVS (typically found at large
distances from their origin at the Galactic Centre) may have lost
nearly all its enhanced metallicity relatively early. Furthermore,
metal-rich HVSs have not been detected \citep{Kollmeier:2011}. We
suggest that if current or future HVSs are found with enhanced
metallicity, it may be due to a relatively recent planetary collision.

\subsection{Bound and Free Planets Around SgrA*} \label{Tidal}

When a binary system is tidally disrupted by the MBH, a HVS may be
produced while the companion star remains in a highly eccentric orbit
around the MBH. If the disruption fails to produce a HVS, both stars
will orbit the MBH. In either case, our simulations show a substantial
probability for a planet to remain bound to a star that is in orbit
around SgrA* (see Table \ref{tab_prob}). It is unclear whether in the
long run such planets will be dynamically stable or eventually fall
into their host stars. However, it is certainly possible that some of
the stars near SgrA* host planetary systems.  Should a star near the
Galactic Center have a planet transit the surface, the change in
magnitude will be $\delta m \sim 0.01$. Near-infrared photometry can
get close to 0.01 mag precision, however extinction and the crowding
of stars make such observations difficult (e.g. \citealt{Scho:10}).

\begin{figure*}
\begin{center}
\begin{tabular}{ccc}
\includegraphics[width=0.5\textwidth]{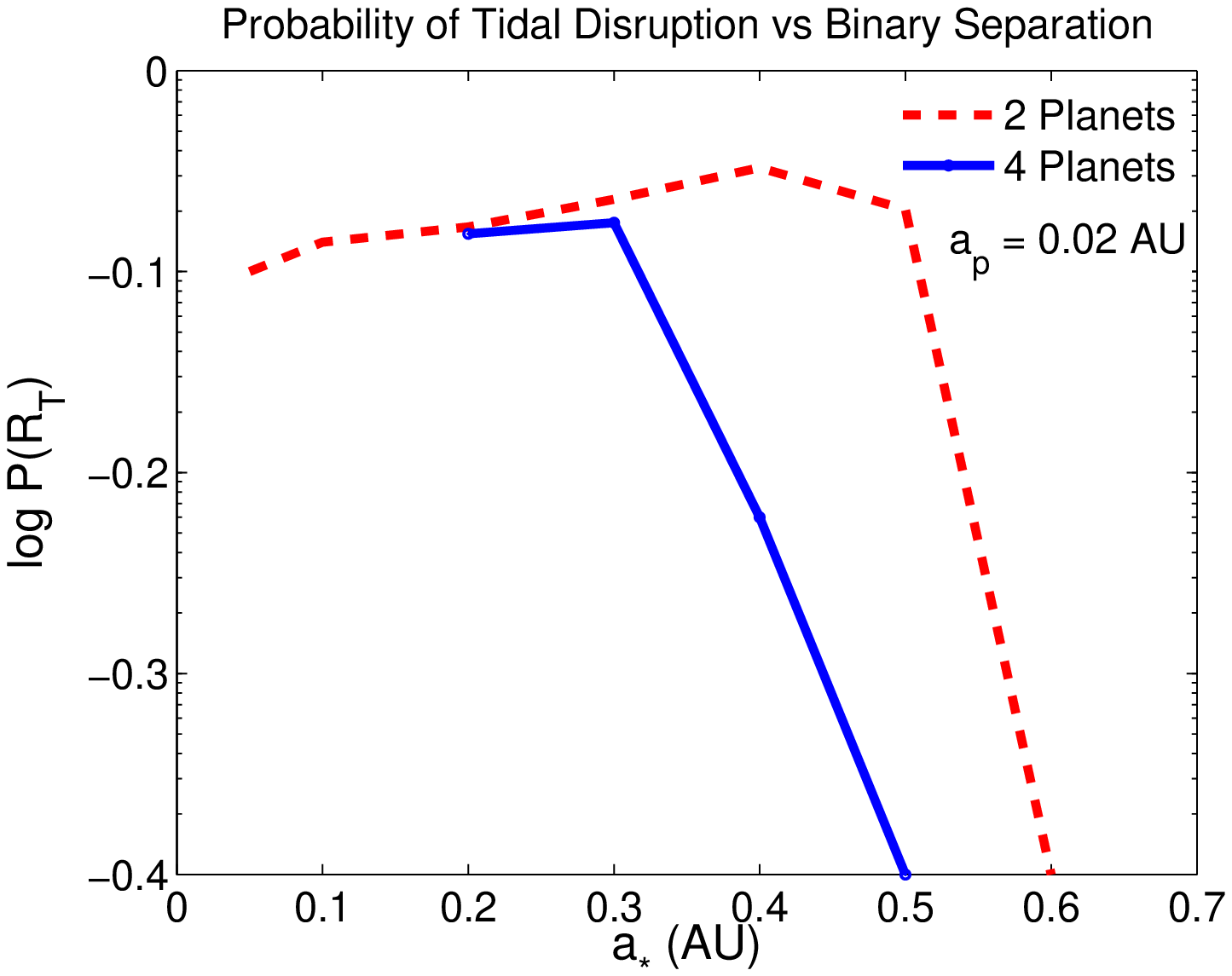}
&
\includegraphics[width=0.5\textwidth]{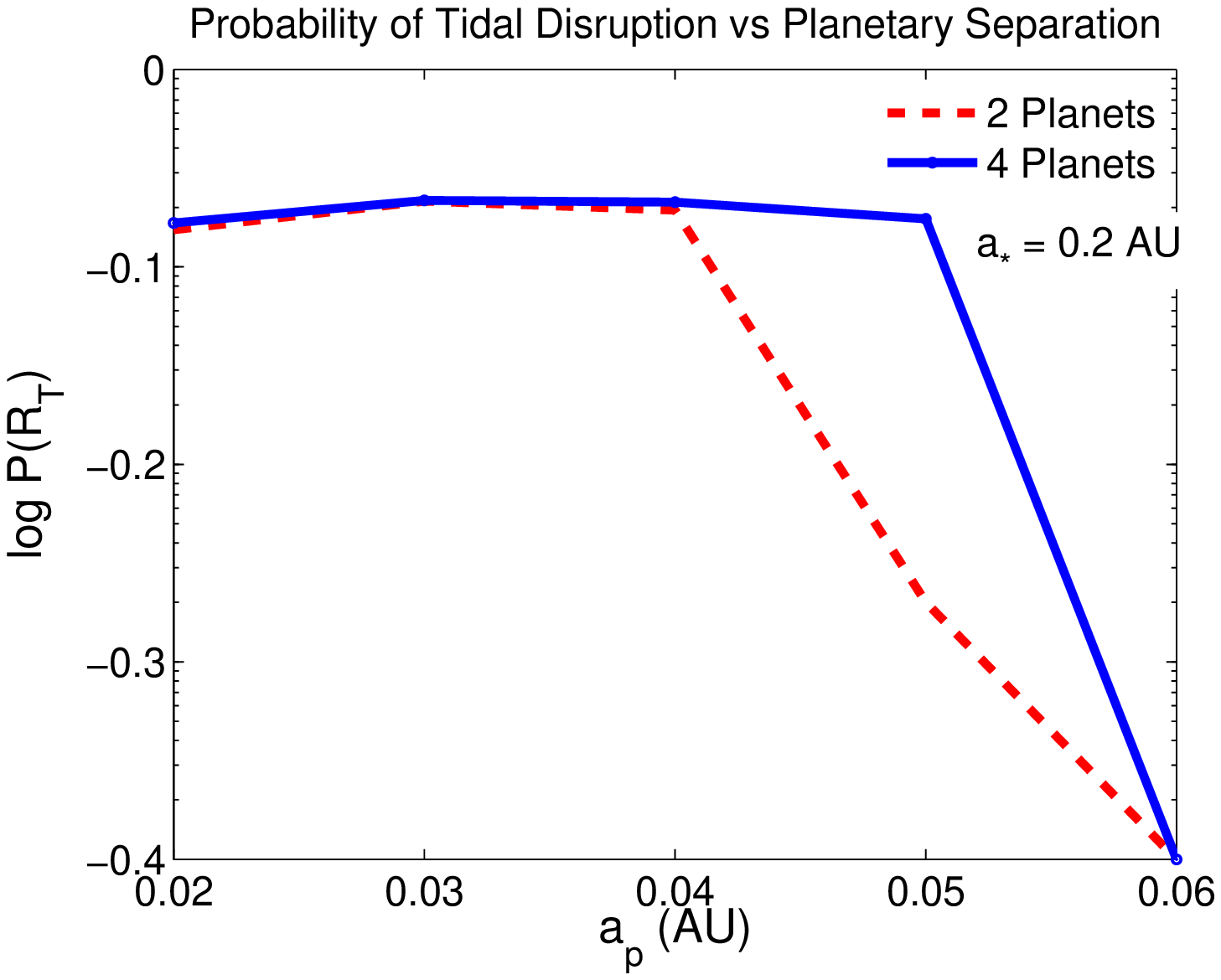}
\\
\end{tabular}
\end{center}
\caption{Probability for a planet freed during the production of a HVS
from a binary star system, to be tidally disrupted by the MBH. {\bf
Left}: Probability as a function of the binary separation. We exclude
simulations with four planets when $a_{\star} < 0.2$ AU due to
instability. {\bf Right:} Probability as a function of planetary
separation. $a_p = 0.02$ AU was the minimum semimajor axis length. For
$a_p = 0.06$ AU the planets either crashed into a star or were
removed, hence the probability is $\sim 0$. The origin of the break
between the simulations with two planets (dashed line) and four
planets (solid line) is due primarily to the fact that for four
planets, the outer planets are easily disrupted with increasing
$a_{\star}$, and are ejected or crash into a star before being tidally
disrupted by the MBH. For increasing $a_p$, both two planet and four
planet simulations have similar stability, however there are more
planets in the latter case and thus the probability does not drop as
quickly.}
\label{fig:prob_disrupt}
\end{figure*}

The probability for at least one planet to 
fall into a highly eccentric orbit around SgrA* is on average $>60$\%
for all our runs. If such a planet passes within the tidal radius of the MBH
\begin{align}
R_{T} \sim R_p\left(\frac{M_{MBH}}{m_p}\right)^{1/3},
\label{eq:tidal}
\end{align}
(where $R_p$ and $m_p$ are respectively the radius and mass of the
planet), the tidal force will disrupt the planet. A star will
similarly be disrupted if it passes too close to its tidal radius, and
might produce an optical flare of $\sim 10^{43}$ erg~s$^{-1}$
\citep{Strubbe-Quataert}. For a planet disrupted by the MBH, the
accretion rate $\dot{M}$ mass would be lower, leading to a luminosity
\begin{align}
L = \epsilon \dot{M}c^2,
\label{eq:eff}
\end{align}
where $\epsilon$ is the radiative efficiency. \citet{Zubovas:11}
suggest that asteroids disrupted by SgrA* may be responsible for
flares with $L \sim 10^{34}-10^{39}$ erg~s$^{-1}$. Furthermore, they
note that a brighter flare ($L\sim10^{41}$ erg~s$^{-1}$) that is
inferred to have occurred $\sim 300$ years ago may be due to the tidal
disruption of a planet. Figure \ref{fig:prob_disrupt} shows the
probability for a planet in orbit around the MBH to be tidally
disrupted. The probability for a system with two planets is shown by
the dashed line, and four planets by the solid line.

\section{Conclusions} \label{Conc}
Our simulations show that in order to produce a HVS with 
planets in orbit, the initial binary separation 
needs to be in the range $a_{\star} = 0.05$--$0.5$ AU, and the planetary
separation in the range $a_p = 0.02$--$0.05$ AU (e.g. ``hot Jupiters''). For such 
parameters there is up to a $\sim 10$\% probability that a 
HVS is produced with orbiting planets (see Table \ref{tab_prob}).
In such scenarios there is a high probability
that at least one planet transits the star (see Figure \ref{fig:transit}).  
For hot Jupiters, the
fractional flux decrement in the light curve, $(R_p/R_{\star})^2 \sim
1$\%, would typically last for hours. Assuming no systematic errors, a
powerful enough telescope can get milli-magnitude photometry
(corresponding to fractional flux sensitivity of 0.1\%) and hence
determine whether or not a HVS has any transiting planets. 
Our examples of planets around HVSs have orbital
periods in the range $0.3$ to $19$ days, requiring observing programs
of this duration. The Hills' mechanism that produces such HVSs also
produces HVPs. Unfortunately, it is extremely difficult to detect a
free HVP. Our simulations indicate that HVPs may achieve speeds
$\sim10^4$ km s$^{-1}$ in rare circumstances.

When a binary system with our given parameters is disrupted, it is
possible that a planet will collide with its host
star. In such instances the surface may be enriched with metals. Such
an enrichment is not believed to last more than a few Myr for massive
stars \citep{Garaud}. The age of HVSs exceed the time necessary for
the excess metallicity in the star's atmosphere to be erased. If
observations show that a HVS has an unusually metal-rich atmosphere,
this may indicate that a planet has relatively recently fallen into
the star's atmosphere.

Planets that are not ejected as HVPs and are not orbiting any HVSs,
may still be detectable as transits around stars orbiting SgrA*. 
Assuming a Jupiter-like planet, the change in magnitude will be 
$\delta m \sim 0.01$. There is also a high probability that 
free-floating planets will be produced. It
may not be possible to ascertain whether a free-floating planet
was originally bound to a binary system that was disrupted. However,
our simulations show that free-floating planets from disrupted systems should be in highly
eccentric orbits around the MBH with an eccentricity $\sim0.96$ or
greater, similar to former companions to HVSs
\citep{Ginsburg:1}. Furthermore, it is also possible for a planet or
star to be tidally disrupted by the MBH, producing a substantial flare
from SgrA*. The luminosity and duration of the flare would indicate
whether it was a planet ($L\sim10^{40}$--$10^{41}$ erg~s$^{-1}$) or star
($L\sim10^{43}$ erg~s$^{-1}$) that fed the MBH
(\citealt{Strubbe-Quataert,Zubovas:11}).

The detection of even one planet around a HVS can shed light on
planetary formation and evolution within the central arcsecond of
SgrA*. In particular, it would support the notion that young stars
near the MBH have close-in planets. It has been suggested that planets
and even asteroids may form in protoplanetary discs around stars
orbiting a MBH \citep{Nayakshin:12}. Recently, a gas cloud was
discovered plunging towards SgrA* \citep{Gillessen:2012}. While it was
suggested that this cloud originated from stellar winds or from a
planetary nebula \citep{Burkert:12}, another theory is that it originated from a
proto-planetary disk around a low-mass star, which implies that
planets may form in the Galactic Centre \citep{Murray-Clay:12}. The
detection of a planet transiting a HVS would lend considerable support
to this interpretation.  If most stars in the Galactic centre form
with planetary systems, then our simulations indicate that there
should be planets in highly eccentric orbits around SgrA*.

\section*{Acknowledgments}

We thank Warren Brown, Smadar Naoz, Dimitar Sasselov, and our 
referee for helpful comments. This work was supported in part by Dartmouth
College, NSF grant AST-0907890, and NASA grants NNX08AL43G and
NNA09DB30A.

\bibliographystyle{mn2e.bst}
\bibliography{HVP.bib}
\bsp

\end{document}